\documentclass[12pt,noshowpacs,nofootinbib,notitlepage,amsmath,amssymb]{revtex4-2}
\usepackage{setspace}
\allowdisplaybreaks
\usepackage{graphicx,color}
\usepackage[colorlinks=true,citecolor=blue,linkcolor=blue,urlcolor=blue]{hyperref}
\usepackage[charter]{mathdesign}
\usepackage{multirow}
\usepackage{enumerate}
\usepackage{array}
\usepackage{booktabs}

\DeclareSymbolFontAlphabet{\mathcal}{symbols}
\DeclareSymbolFont{symbols}{OMS}{xmdcmsy}{m}{n}
\DeclareSymbolFont{largesymbols}{OMX}{cmex}{m}{n}
\SetSymbolFont{symbols}{bold}{OMS}{xmdcmsy}{b}{n}

\begin{document}  
\title{\color{blue}\Large Running couplings and unitarity\\in a 4-derivative scalar field theory}

\author{Bob Holdom}
\email{bob.holdom@utoronto.ca}
\affiliation{Department of Physics, University of Toronto, Toronto, Ontario, Canada  M5S 1A7}
\begin{abstract}
We obtain the $\beta$-functions for the two dimensionless couplings of a 4d renormalizable scalar field theory with cubic and quartic 4-derivative interactions. Both couplings can be asymptotically free in the UV, and in some cases also in the IR. This theory illustrates the meaning of unitarity in the presence of a negative norm state. A perturbative calculation that accounts for the new minus signs shows that the optical theorem is identically satisfied. These minus signs also enter a discussion of tree-level scattering. For a certain setup involving colliding beams of particles we find even more intricate cancellations and quite normal behaviour at high energies. The $\beta$-functions for the Stuckelberg gauged version of the theory are also obtained.
\end{abstract}
\maketitle 

\section{Introduction}
We consider the following real scalar field theory in four spacetime dimensions, with four derivatives appearing in both the kinetic term and the cubic and quartic interaction terms,
\begin{align}
{\cal L}_1=\frac{1}{2}\partial_\mu\phi(\Box+m^2)\partial^\mu\phi+\lambda_3(\partial_\mu\phi\partial^\mu\phi)\,\Box\phi+\lambda_4(\partial_\mu\phi\partial^\mu\phi)^2.\label{e1}
\end{align}
The scalar field $\phi=\phi(x)$ is dimensionless, as are the couplings $\lambda_3$ and $\lambda_4$. We shall allow these couplings to have either sign. The mass term breaks a classical scale invariance. At low enough energies only the mass term survives and the field can be re-scaled to absorb the mass, leaving a free and massless theory. Our interest is then in the UV behaviour of the theory. Many of our results will be obtained by calculating with a finite mass and then taking the $m\to0$ limit at the end. These mass independent results will exhibit the manner in which the scale invariance is broken quantum mechanically. We do not claim that this $m\to0$ limit should be taken as the definition of the theory where $m$ is identically zero to begin with. That theory has problems with its definition and cannot be handled directly by standard methods. The theory in (\ref{e1}) can be.

This theory also exhibits a simple shift symmetry $\phi(x)\to\phi(x)+c$. A term has a shift symmetry if a derivative acts on every factor of $\phi(x)$, or if not, the term is equivalent to one that is by integration by parts. Since the field is dimensionless the shift symmetry excludes an infinite number of terms with dimensionless couplings. If the $\phi$ field couples to any other field then the shift symmetry must also be preserved by that coupling. The shift symmetry, along with the four derivative kinetic term and the lack of couplings with inverse mass dimensions, ensures that the theory is renormalizable. The two simpler theories, with $\lambda_3=0$ or $\lambda_4=0$ respectively, are also renormalizable, and sometimes our focus will be on one of those theories.

Our first goal is to study the renormalization of (\ref{e1}) and to obtain the one-loop $\beta$-functions for the couplings $\lambda_3$ and $\lambda_4$. We will display the flow diagram of the running couplings that shows their behaviour in both the UV and the IR, where by IR we mean scales that are still above the arbitrarily small mass $m$. Perhaps most notable is that there is a family of trajectories where both couplings are asymptotically free in both the UV and the IR. Our other goals are to study the unitarity of the theory and the behaviour of scattering at high energies. We comment on the Stuckelberg extension of the theory at the end.

A higher derivative scalar field theory has been extensively studied in another context \cite{ant,Elizalde:1994sn,Antoniadis:1994ij}. This is the theory of the conformal factor of the metric as extracted from an effective action related to the conformal anomaly, which in turn is generated by loops of light particles. A simple form of such a Lagrangian using conventional notation is
\begin{align}
{\cal L}_{\rm conf}=-\theta^2(\Box \sigma)^2-\zeta[2(\partial\sigma)^2\Box\sigma+(\partial\sigma)^4]+\gamma e^{2\sigma}(\partial\sigma)^2-\lambda e^{4\sigma}.
\label{e19}\end{align}
The corresponding action has a global conformal symmetry, and this is what relates the coefficients in the second term and brings in the exponential factors in the last two terms, which are the mass and cosmological constant terms. The interest in (\ref{e19}) is usually driven by its effective description of possible infrared physics whereas our interest in (\ref{e1}) is in the UV. Calculations of $\beta$-functions for (\ref{e19}) usually do not treat renormalization of $(\Box \sigma)^2$ as a wave function renormalization, and so give different results. $\beta$-functions of the more standard type were given in \cite{Elizalde:1994sn}, but there are various minus sign and factor of 2 deviations from our results. Very recent results for the renormalization of (\ref{e19}) have been given in \cite{Silva:2023lts}, without giving $\beta$-functions, and our results agree where they overlap. 

Other recent studies of higher derivative scalar theories have appeared in various contexts \cite{Safari:2021ocb,Buccio:2022egr,Tseytlin:2022flu}, but without considering both the $\lambda_3$ and $\lambda_4$ couplings simultaneously. There is interest in these theories even with an assumed lack of unitarity due to the existence of a negative norm state. But there is more to the unitarity story than this. As far as we are aware, the authors of \cite{Abe:2018rwb} were the first to look more closely at the optical theorem in higher derivative scalar theories and its relation to renormalizability. (One of their criteria for renormalizability appears to be equivalent to requiring the shift symmetry to apply only to terms with dimensionless couplings.) These authors derive an inequality from the optical theorem at weak coupling and show by example that it is satisfied only when the theory is renormalizable, with their renormalizable example being the $\lambda_3=0$ theory. The authors refer to this as S-matrix unitarity, the fact that $SS^\dagger=1$ can still hold, even though it no longer implies the standard bound on the size of amplitudes at high energies.

Another motivation for the study of (\ref{e1}) is its similarity to quantum quadratic gravity. The latter is also a renormalizable quantum field theory \cite{stelle} with four derivatives in all terms except for a mass term with two derivatives. The mass term is the Einstein term and the two four-derivative interaction terms are curvature-squared terms with dimensionless couplings. The mass term again breaks a classical scale invariance, while the coordinate invariance of quadratic gravity is the analog of the shift symmetry of (\ref{e1}). Quadratic gravity is a significantly more complex theory, and so the theory in (\ref{e1}) is thus a useful proxy for quadratic gravity where we can much more easily obtain loop amplitudes, and where we can in particular investigate the optical theorem.

With the completeness relation
\begin{align}
\mathbb{1}=\sum_X\int d\Pi_X \frac{|X\rangle \langle X|}{\langle X|X \rangle}
\label{e24}\end{align}
inserted on the RHS of $-i(T-T^\dagger)=TT^\dagger$, where $S=1+iT$, we see that there are new minus signs on the RHS of the optical theorem when a state $|X\rangle$ has a negative norm. These signs produce the cancellations among the terms on the RHS as observed by \cite{Abe:2018rwb}. By focusing on the inequality satisfied by the RHS, those authors avoided calculating the one-loop amplitude appearing on the LHS of the optical theorem. In Section \ref{s3} we shall use the $\lambda_3=0$ theory to calculate both the LHS and RHS to show that the optical theorem at order $\lambda_4^2$ is identically satisfied, both in the $m\to0$ limit and for finite $m$. We can refer to the theory as having unitarity without positivity.

We then argue that the structure of the theory puts constraints on the scattering experiments that can actually be physically realized. In Section \ref{s4} we calculate the lowest order event rates in such scattering experiments and find that they can have high energy behaviour similar to theories with standard unitarity bounds. This is due to even more intricate cancellations brought about by the lack of positivity among the terms summed in an inclusive process. We have also found these types of cancellations for the scattering of gravitational degrees of freedom in \cite{Holdom:2021hlo} and for photon-photon scattering in \cite{Holdom:2021oii}, both in the context of quantum quadratic gravity.

\section{Renormalization and running couplings}
With the following transformations,
\begin{align}
\phi&=\psi_1-\psi_2,\label{e3}\\
\psi_1&=\frac{1}{m^2}(\Box+m^2)\phi,\label{e4}\\
\psi_2&=\frac{1}{m^2}\Box\phi,\label{e5}
\end{align}
Lagrangian ${\cal L}_1$ can be converted to Lagrangian ${\cal L}_2$,
\begin{align}
{\cal L}_2&=-\frac{m^2}{2}\psi_1\Box\psi_1+\frac{m^2}{2}\psi_2(\Box+m^2)\psi_2\nonumber\\&+\lambda_3\left(\partial\psi_1-\partial\psi_2\right)^2(\Box\psi_1-\Box\psi_2)\label{e2}\\&+\lambda_4\left(\partial\psi_1-\partial\psi_2\right)^4.\nonumber
\end{align}
${\cal L}_2$ is familiar for having the wrong sign kinetic term for $\psi_2$. We have chosen to keep the $\psi_1$ and $\psi_2$ fields dimensionless, since it is not difficult to account for the $m^2$ normalization of the kinetic terms in calculations. Note that the shift symmetry $\phi\to\phi+c$ corresponds to $\psi_1\to\psi_1+c$ and $\psi_2\to\psi_2$.\footnote{That these two Lagrangians are the same can be deduced by substituting (\ref{e3}) into the quadratic term in (\ref{e1}), taking the difference with the quadratic terms in (\ref{e2}), and then substituting (\ref{e4}) and (\ref{e5}) into the result to find zero after an integration by parts.}

Lagrangian ${\cal L}_1$ can be taken to define a propagator of a two pole form,
\begin{align}
-\frac{i}{(k^2+i\epsilon)(k^2-m^2+i\epsilon)}\label{e8}
.\end{align}
This gives Feynman diagrams where the number of propagators is effectively doubled, where each pair of propagators with the same momentum can be interpreted to bring in an additional zero-momentum vertex. The diagram can thus be evaluated by standard methods. But when the effective number of propagators grows to six or more then standard methods are not so well developed and this approach quickly becomes cumbersome.

For the second Lagrangian ${\cal L}_2$, the number of fields is doubled, and so each internal line can be one of the following two propagators,
\begin{align}
\frac{1}{m^2}\,\frac{i}{k^2+i\epsilon},\quad-\frac{1}{m^2}\,\frac{i}{k^2-m^2+i\epsilon}.
\end{align}
Only the first has the normal sign. The number of diagrams quickly grows with the number of internal lines, but standard methods can now more easily deal with, say, four propagators in a loop, rather than eight that would occur from ${\cal L}_1$.

We can suppose that either of these Lagrangians are composed of renormalized fields and renormalized couplings $m$, $\lambda_3$ and $\lambda_4$. These renormalized quantities are related to bare quantities via the renormalization constants $\phi_0=Z_\phi^\frac{1}{2}\phi$, $m_0^2=Z_m m^2$, $\lambda_{30}=Z_3\lambda_3$ and $\lambda_{40}=Z_4\lambda_4$. A counterterm Lagrangian is constructed using these renormalization constants such that the total Lagrangian yields finite results at one-loop. The respective counterterm Lagrangians are
\begin{align}
{\cal L}_1^{\rm c.t.}&=\frac{1}{2}\partial\phi\!\left[(Z_\phi-1)\Box+(Z_\phi Z_m-1)m^2\right]\!\partial\phi\nonumber\\
&+(Z_\phi^\frac{3}{2} Z_3-1)\lambda_3(\partial\phi)^2\Box\phi\label{e20}\\
&+(Z_\phi^2 Z_4-1)\lambda_4(\partial\phi)^4,\nonumber
\end{align}
\begin{align}
{\cal L}_2^{\rm c.t.}&=-\frac{m^2}{2}\psi_1(Z_\phi Z_m-1)\Box\psi_1+\frac{m^2}{2}\psi_2\!\left[(Z_\phi Z_m-1)\Box+(Z_\phi Z_m^2-1)m^2\right]\!\psi_2\nonumber\\
&+(Z_\phi^\frac{3}{2} Z_3-1)\lambda_3\left(\partial\psi_1-\partial\psi_2\right)^2(\Box\psi_1-\Box\psi_2)\label{e14}\\
&+(Z_\phi^2 Z_4-1)\lambda_4\left(\partial\psi_1-\partial\psi_2\right)^4.
\nonumber\end{align}
The total Lagrangians ${\cal L}_1+{\cal L}_1^{\rm c.t.}$ and ${\cal L}_2+{\cal L}_2^{\rm c.t.}$ when written in terms of bare quantities take the same form as the original Lagrangians ${\cal L}_1$ and ${\cal L}_2$. We have used $Z_{\psi_1}=Z_{\psi_2}=Z_\phi$.

\begin{figure}[h]
\begin{center}\vspace{-0ex}
\includegraphics[width=1\textwidth]{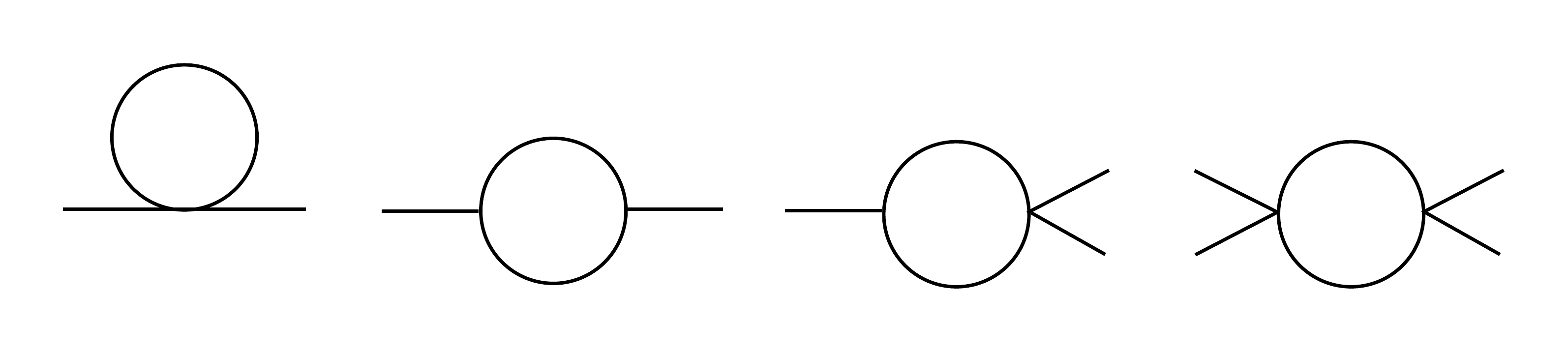}
\vspace{-5ex}\caption{Divergent diagrams.}
\label{f3}
\end{center}
\end{figure}

Either ${\cal L}_1$ or ${\cal L}_2$ can be used to extract the various renormalization constants. Using dimensional regularization with
\begin{align}
\frac{2}{\varepsilon}=\frac{2}{4-d}-\gamma_E+\ln(4\pi),
\end{align}
we find the following results\footnote{Our calculations use a combination of Maple and Package-X \cite{Patel:2015tea} for Mathematica.}
\begin{align}
Z_\phi Z_m-1&=-\frac{3\lambda_4}{4\pi^2}\frac{1}{\varepsilon},\label{e16}\\
Z_\phi -1&=\frac{5\lambda_3^2}{8\pi^2}\frac{1}{\varepsilon},\label{e15}\\
Z_\phi^\frac{3}{2} Z_3-1&=-\frac{5\lambda_4}{4\pi^2}\frac{1}{\varepsilon},\label{e18}\\
Z_\phi^2 Z_4-1&=-\frac{5\lambda_4}{4\pi^2}\frac{1}{\varepsilon}.\label{e17}
\end{align}
The one-loop one-particle-irreducible diagrams that produce these $\varepsilon$-poles are those that involve one or two propagators, and the four diagrams corresponding to (\ref{e16}), (\ref{e15}), (\ref{e18}), (\ref{e17}) respectively are shown in Fig.~\ref{f3}.\footnote{The use of ${\cal L}_1$ to calculate these diagrams is somewhat simpler, and if ${\cal L}_2$ is used to extract $Z_\phi$ and $Z_m$, the one-loop corrections that emerge are in the form of the kinetic terms in ${\cal L}_1$ after inserting $\phi=\psi_1-\psi_2$.}
\vspace{-0ex}

\begin{figure}[h]
\begin{center}\vspace{-0ex}
\includegraphics[width=.9\textwidth]{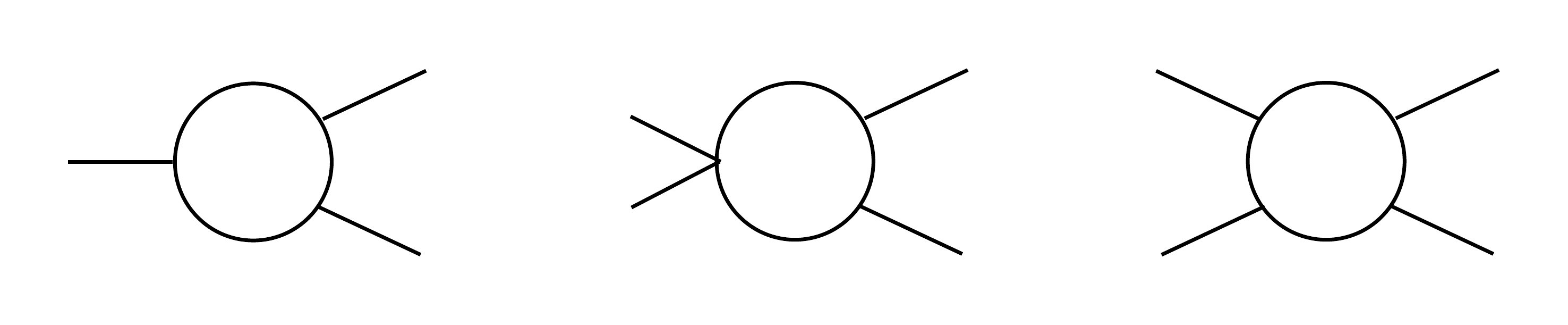}
\vspace{-0ex}\caption{Finite diagrams.}
\label{f4}
\end{center}
\end{figure}

There are also 3 and 4-point one-particle-irreducible diagrams that have more than two propagators, as shown in Fig.~\ref{f4}. Power counting arguments suggest that these diagrams should also have log divergences, similar to the diagrams in Fig.~\ref{f3}. But this is not the case; we find that the diagrams in Fig.~\ref{f4} are finite by explicit calculation using ${\cal L}_2$. (For the first diagram in Fig.~\ref{f4} this was noted in \cite{Safari:2021ocb}.) Why this is so is related to renormalizability. The diagrams in Fig.~\ref{f4} can be converted to 5, 6, 7 or 8-point functions by converting cubic vertices to quartic vertices. Power counting suggests that those diagrams also have log divergences. But there are no counter-terms with these numbers of fields, due to the shift symmetry, and so the absence of divergences for such diagrams must be due to the shift symmetry. It thus appears that the diagrams in Fig.~\ref{f4} are finite for this reason as well.
\vspace{-0ex}

The nature of the divergence in the first diagram in Fig.~\ref{f3}, which renormalizes the $m^2\partial\phi\partial\phi$ term, is different from the other three. Other than the factor of $p^2m^2$, there is no further momentum dependence and thus no scale dependence from this tadpole-type diagram. Effectively (\ref{e16}) should be replaced by $Z_\phi Z_m=1$ at order $1/\varepsilon$. If we define the anomalous dimensions as $\gamma=\frac{1}{2}dZ_\phi/d\ln\mu$ and $\gamma_m=\frac{1}{2}dZ_m/d\ln\mu$ then we have
\begin{align}
\gamma=-\gamma_m=\frac{5\lambda_3^2}{16\pi^2}.
\end{align}
The running of the mass is due entirely to the wave function renormalization, due to the second diagram in Fig.~\ref{f3}, and the mass term as a whole has no scale dependence.

\begin{figure}[h]
\begin{center}
\includegraphics[width=.8\textwidth]{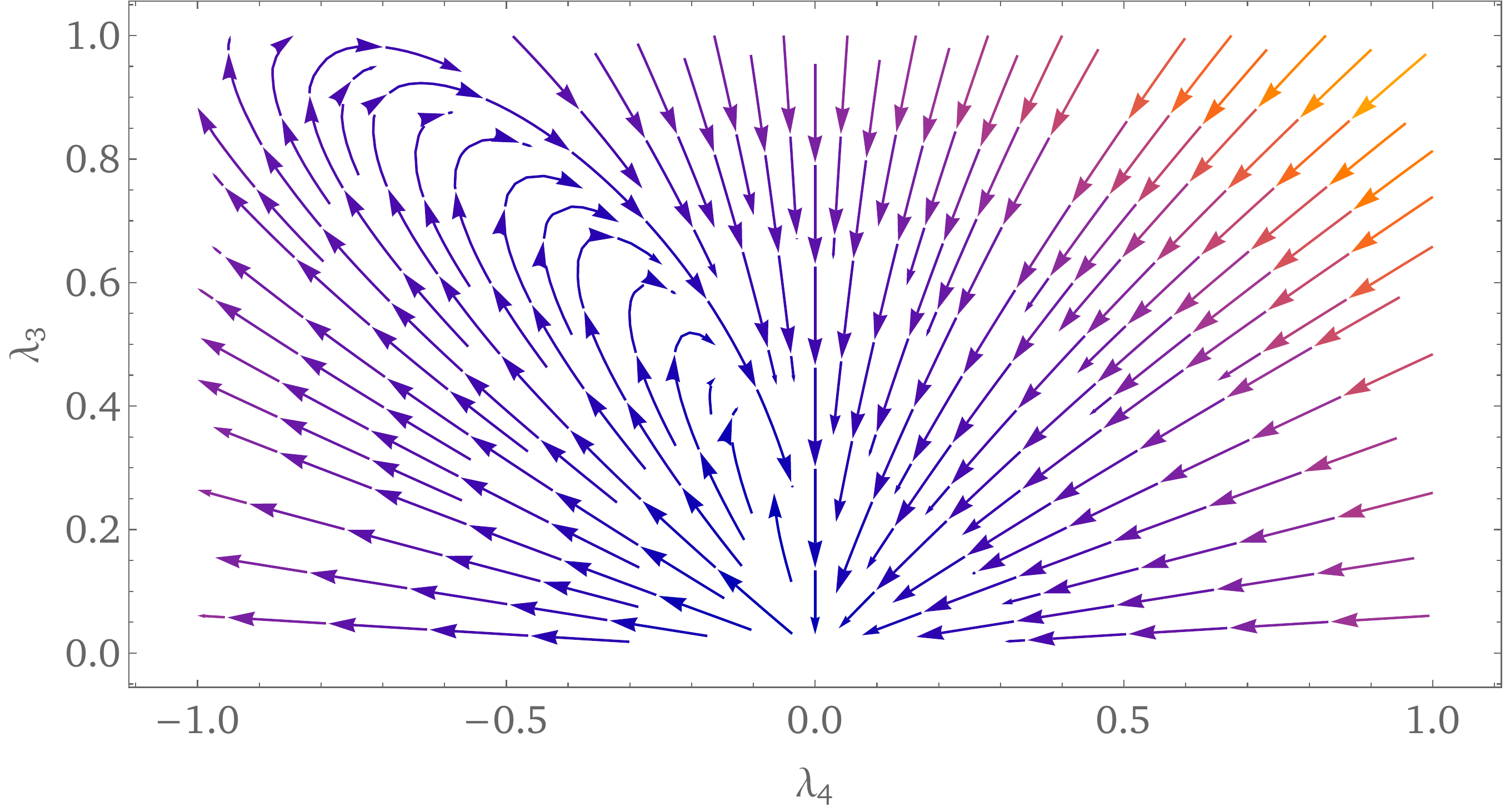}
\caption{Flow diagram for the two couplings where the arrows point towards the UV. The extension to negative $\lambda_3$ is obtained by flipping the above diagram around the line $\lambda_3=0$.}
\label{f2}
\end{center}
\end{figure}

The other three diagrams in Fig.~\ref{f3} do introduce nontrivial scale dependence. The $\beta$-functions in $d=4$ dimensions are given by the residues of the $1/(4-d)$ poles in $Z_3 \lambda_3$ and $Z_4 \lambda_4$ respectively,
\begin{align}
\frac{d\lambda_3}{d\ln\mu}&=-\frac{5}{4\pi^2}(\lambda_4\lambda_3+\frac{3}{4}\lambda_3^3),\label{e7}\\\frac{d\lambda_4}{d\ln\mu}&=-\frac{5}{4\pi^2}(\lambda_4^2+\lambda_4\lambda_3^2).\label{e6}
\end{align}
The second term in each $\beta$-function is due to the wave function renormalization. Clearly both couplings are asymptotically free in the UV when both are positive. For general signs, we show the flow diagram for the two couplings in Fig.~(\ref{f2}). Interesting behaviour occurs in the quadrants where $\lambda_4<0$. Then there are trajectories where both couplings are asymptotically free in both the UV and the IR, such that $|\lambda_3|$ and $|\lambda_4|$ reach maximum values at intermediate scales. The decrease of $|\lambda_3|$ and $|\lambda_4|$ in the IR will continue down to the scale of the mass, below which the theory is free.

Our results for the $\beta$-functions depend on the normalizations and signs in our definition of the theory in (\ref{e1}). For example it may be common to reverse the signs of the interaction terms, in which case (\ref{e7}, \ref{e6}) are transformed with $\lambda_3\to-\lambda_3$ and $\lambda_4\to-\lambda_4$. The resulting flow diagram is obtained from the one in Fig.~(\ref{f2}) by flipping it around the line $\lambda_4=0$. In \cite{Safari:2021ocb} the $\lambda_3=0$ and $\lambda_4=0$ theories are treated separately, and so they obtain the $\lambda_3^3$ term in (\ref{e7}) and the $\lambda_4^2$ term in (\ref{e6}) (the latter is also obtained in \cite{Tseytlin:2022flu}). Our results agree.

With dimensional regularization we can reduce the number of spacetime dimensions $d$ and look for $1/(2-d)$ poles, which would correspond to quadratic divergences in four dimensions. We find a $1/(2-d)$ pole only in one case, in the tadpole diagram associated with the mass renormalization. The other diagrams in Fig.~\ref{f3} are finite in two dimensions.

\section{Unitarity and the optical theorem}\label{s3}
First we give the full momentum dependent amplitudes corresponding to the last three diagrams in Fig.~\ref{f3}, after taking the $m/E\to0$ limit. No infrared divergences are encountered in this limit. The Feynman amplitudes are a factor of $i$ times the following two, three and four-point functions. The $1/\varepsilon$ poles appearing here are cancelled by the Lagrangian counterterms as determined by (\ref{e15}), (\ref{e18}), (\ref{e17}) respectively.\footnote{The overall sign of these results depends on the number of $\psi_2$ fields, as is the case for the tree-level vertices. The results presented assume an even number. This is also the case of interest for the optical theorem, since elastic scattering involves the same particles in the initial and final states.}
\begin{align}
\Sigma(p)&=\frac{\lambda_3^2}{(4\pi)^2}(p^2)^2\left(\frac{10}{\varepsilon}+3+5\log\left(-\frac{\mu^2}{p^2}\right)\right)\\
\Gamma_3(p,q)&=\frac{\lambda_3\lambda_4}{(4\pi)^2}\left((\frac{80}{\varepsilon}+\frac{112}{3})((pq)^2-p^2q^2)\right.\\
& -\frac{8}{3}[(7 p^2 q^2+6 (pq)p^2 - (pq)^2) \log(-\frac{\mu^2}{p^2}) + ( 7 p^2 q^2+6 (pq) q^2-(pq)^2 ) \log(-\frac{\mu^2}{q^2})\nonumber \\&\left.+(p^2 q^2 - 6 (pq)(p^2+ q^2)  - 13 (pq)^2)  \log(-\frac{\mu^2}{(p+q)^2})]\right)\nonumber\\
\Gamma_4(s,t)&=\frac{\lambda_4^2}{(4\pi)^2}\left((\frac{80}{\varepsilon}+\frac{112}{3})(s^2+st+t^2)+i\frac{8}{3}\pi\,(7s^2+st+t^2)\right.\label{e21}\\&
\left.+\frac{8}{3} [(7 s^2 + s t + t^2) \log(\frac{\mu^2}{s}) + (s^2 + s t + 7 t^2) \log(-\frac{\mu^2}{t}) + (7 s^2 + 13 s t + 7 t^2) \log(\frac{\mu^2}{s + t})]\right)\nonumber
\end{align}

Our focus shall be on the optical theorem for two-to-two scattering where we restrict to the $\lambda_3=0$ theory, and for which the four-point function $\Gamma_4(s,t)$ is of interest. We have assumed that the four external particles are on-shell, and so the $m/E\to0$ limit produces a four-point function that is a function of the Mandelstam variables $s$ and $t$.\footnote{A similar result is obtained in \cite{Tseytlin:2022flu}.} It is evaluated in the physical region where $s>0,t<0,s+t>0$ such that the imaginary part is displayed explicitly. For elastic scattering in the forward direction we are interested in $\Gamma_4(s,0)$; twice its imaginary part is the $s$ cut discontinuity that is related by the optical theorem to an inclusive cross section (with the initial state normalization factor omitted). We see from (\ref{e21}) that this discontinuity is $7s^2/3\pi$.

Before checking the optical theorem we first mention some factors appearing in an exclusive cross section. Most importantly the $\psi_2$ state is a negative norm state, and this results in a factor $(-1)^{n_2}$ where $n_2$ is the number of $\psi_2$'s in the initial and final states.\footnote{A cross section is related to a probability and this involves dividing by the norms of the states involved, similar to the appearance of the norm in the completeness relation (\ref{e24}).} The presence of the $m^2$ factors in the kinetic terms of ${\cal L}_2$ also implies an additional normalization factor of $1/m^2$ for each $\psi_1$ or $\psi_2$ state. These factors would seem to make the $m\to0$ limit problematic, but this is not the case for the optical theorem and for our discussion in the next section.  We can also comment on the LSZ reduction formulas that extract scattering amplitudes from the residues of the poles of Greens functions. For Lagrangian ${\cal L}_1$ each external line is associated with two poles, and so the residues of the various products of poles give the scattering amplitudes. For Lagrangian ${\cal L}_2$ the reduction formula is more standard, and indeed, the definition of $\psi_1$ and $\psi_2$ in (\ref{e4}) and (\ref{e5}) can be seen to leave just one pole for each external line.

The RHS of the optical theorem involves a sum of squares of tree-level, on-shell, two-to-two amplitudes generated by the quartic interaction. These tree-level amplitudes are
\begin{align}
i(-1)^{n_2}8\lambda_4[(p_1p_2)(p_3p_4)+(p_1p_3)(p_2p_4)+(p_1p_4)(p_2p_3)].
\end{align}
The $(-1)^{n_2}$ amplitude factor here is not to be confused with the same factor in an exclusive cross section. The squares of these amplitudes are summed over the possible final states for a fixed initial state, with the norm and normalization factors included for the possible final states, but not for the initial state. (The RHS of the optical theorem differs from a cross section by this initial state factor.) We employ on-shell kinematics for all particles and express each amplitude in terms of $s$ and $m$ and the scattering angle in the centre-of-momentum frame. The angular integration of the sum of squares is done with a factor of $\frac{1}{2}$ to avoid double counting, after which we finally let $m/E\to0$. We find that the result is $7s^2/3\pi$ and so the optical theorem is satisfied. Any of the initial states, $\psi_1\psi_1$, $\psi_2\psi_2$ or $\psi_1\psi_2$ gives the same result.

We can also consider the finite mass version of the optical theorem, and for this we restrict to the $\psi_1\psi_1$ initial state. For the LHS we evaluate the fourth diagram of Fig.~\ref{f3} without taking the $m/E\to0$ limit and find that the discontinuity across the $s$ cut is
\begin{align}
\frac{\lambda_4^2}{30 m^4 \pi} &\left(21 s^4 \,\theta(s) - 
   \frac{2}{s} (s-m^2 )^3 (21 s^2 + 8 m^2 s + m^4) \,\theta(s-m^2)\right.\label{e22}\\&\left.+\; 
   s \sqrt{s (s-4 m^2)} (21 s^2 - 68 m^2 s +56 m^4)\, \theta(s-4 m^2) \right)\nonumber
.\end{align}
For the RHS of the optical theorem we just avoid taking the $m/E\to0$ limit in the previous evaluation of the RHS. The result is again (\ref{e22}) and so this confirms the optical theorem for any $m$. Due to the negative sign of the $\theta(s-m^2)$ term, at large $s$ the cancellations produce $s^2$ behaviour rather than the naive $s^4/m^4$ behaviour. This minus sign is due to a negative propagator when the LHS is calculated, and it is due to a negative norm when the RHS is calculated. So it is not a surprise that the optical theorem is satisfied. The theory is displaying unitarity without positivity. This more accurate description allows us focus on the actual nonstandard property of the theory, which is the lack of positivity.

The RHS of the optical theorem becomes a cross section when multiplying by the initial state factor $\propto1/(sm^4)$. It is this factor that introduces a minus sign for the $\psi_1\psi_2$ initial state, meaning that this cross section is negative. But $\psi_2$ does not describe an asymptotic state and so this initial state cannot be treated in isolation. The massive $\psi_2$ particle is unstable, decaying into $2\psi_1$ or $3\psi_1$ or any other light particles to which it couples. The proper time between the original production of the $\psi_2$ and its scattering must be less than the $\psi_2$ lifetime. There is another minus sign built into the $\psi_2$ production cross section, and so there are two minus signs, in the production and scattering cross sections respectively, that cancel in the product. A positive probability is associated with the complete $n_2$-even process.

We shall instead turn to high energies and short time scales so that both mass and decay effects are small. Within this context we can consider a standard scattering experiment involving the collision of two beams of particles. The two beams should be composed of nearly equal numbers of $\psi_1$ and $\psi_2$, since the $\psi_2$ or $\psi_1$ particles are produced through couplings that always involves the combination $\psi_1-\psi_2$ and so the two particles are equally likely to be produced. With this setup it is a sum of probabilities that is positive, and it is this that is related to certain physical observables in the high energy theory. We develop this picture in the next section.

If it was possible to have pure $\psi_1$ beams, for example by utilizing long enough times for the initial beams to be depleted in $\psi_2$'s, due to their decay, then the cross section would behave like $s/m^4$ for large $s$. This violation of an upper bound, that is normally attributed to unitarity, is allowed when unitarity comes without positivity \cite{Abe:2018rwb}. But there is a caveat regarding this picture of on-shell scattering at large $s$. It may be that the analog of parton showers occurs in both the initial and final states, which would imply that the actual hard scattering occurs between highly virtual excitations of the $\psi_1$ and $\psi_2$ fields. Then on-shell particles and negative norm states are not actually involved in the hard scattering process. So it is still not certain that the $s/m^4$ behaviour can be realized. In any case, the on-shell picture with suitably inclusive differential cross sections can still provide a useful dual description of the off-shell scattering, as it does in perturbative QCD. It is the inclusive, rather than the exclusive, on-shell cross sections that are meaningful in this context. This view of the on-shell description was taken in \cite{Holdom:2021hlo} and the discussion of the next section can be viewed in this light.

\section{A scattering experiment}\label{s4}
The sum or average over $\psi_1$ and $\psi_2$ in the initial state is trickier to deal with than an averaging over, for example, polarizations. As before we calculate for finite $m$ first before taking the $m/E\to0$ limit. Since intricate cancellations take place in this limit, we need to properly account for all mass effects before the limit is taken. We shall introduce a luminosity factor to better describe the initial state, since this factor is another source of mass dependence. We assume short time scales so that the effect of $\psi_2$ decay can be ignored.

Consider the collision of two finite beams of particles, each containing both types of particles.  A choice of initial state $i=(A,B)$ corresponds to considering the collision of particles of type $A$ in one beam with particles of type $B$ in the other beam. And for a choice $i$ we can consider the integrated luminosity ${\cal L}^{\rm int}_i$ such that its product with the cross section $\sigma_{i\to f}$ gives the number of scattering events from initial state $i$ to final state $f$. The integrated luminosity is an integral of the Lorentz-invariant luminosity density $S$, \cite{furman}
\begin{align}
{\cal L}^{\rm int}_i=\int dtd^3\textbf{x} \,S,\quad S=\sqrt{(j_A j_B)^2-j_A^2 j_B^2}.
\end{align}
The 4-vector $j^\mu$ is a particle current density. $j^\mu$ is proportional to $p^\mu$, and in particular $j^\mu=(\rho/E)p^\mu$ where $\rho$ is a particle density. Thus $\rho/E$ is a Lorentz invariant constant and we can write
\begin{align}
S&=\frac{\rho_A\rho_B}{E_A E_B}\sqrt{(p_A p_B)^2-m_A^2 m_B^2}.\label{e9}
\end{align}
The actual values of $\rho_A$ and $\rho_B$ are functions of the respective momenta $p_A$ and $p_B$ of the particles $A$ and $B$ occurring in the beams.

Now let us go to the centre-of-momentum frame where
\begin{align}
{\cal L}^{\rm int}_i\sigma_{i\to f}&=F_{AB}\frac{\sqrt{s}p_i}{E_A E_B}\sigma_{i\to f},\label{e11}\\
F_{AB}&=\int dtd^3\textbf{x} \,\rho_A\rho_B.\nonumber
\end{align}
We have used $\sqrt{(p^\mu_A p_{B\mu})^2-m_A^2 m_B^2}=\sqrt{s}p_i$, where $p_i$ is the magnitude of the equal and opposite 3-momenta describing the initial state in this frame. $F_{AB}$ describes the beams and has dimensions of inverse area. Also in this frame the exclusive differential cross section is
\begin{align}
\frac{d\sigma_{i\to j}}{d\Omega}=\frac{(-1)^{n_2}}{m^8}\frac{1}{64\pi^2 s}\frac{p_f}{p_i}|{\cal M}_{fi}(s,\theta)|^2,\label{e12}
\end{align}
where $\Omega=(\theta,\phi)$ characterizes the direction of a final state 3-momentum, of magnitude $p_f$, relative to the direction of an initial state 3-momentum. Note that $p_i$, $p_f$, $E_A$, $E_B$ are all determined from the value of $s$ and the masses and from the constraint that all particles are on shell. We have mentioned the origin of the $(-1)^{n_2}/m^8$ factor above.

It is the inclusive differential cross section, the sum of the exclusive ones, that can have a physical meaning. In particular the combination of (\ref{e11}) and (\ref{e12}) shows that the total number of scattering events in differential form is given by
\begin{align}
\frac{dN}{d\Omega}=\sum_{i,f}\frac{(-1)^{n_2}}{64\pi^2}F_{AB}\frac{1}{E_A E_B}\frac{p_f}{\sqrt{s}}\frac{1}{m^8}|{\cal M}_{fi}(s,\theta)|^2.\label{e10}
\end{align}
The sum over $i$ and $f$ sums over all combinations of $\psi_1$ and $\psi_2$ in the initial and final states that are kinematically allowed by the value of $s$ being considered. Although we are only calculating this at tree level, the imposition of the on-shell constraints produces a lengthy expression. Cancellations occur in the $m/E\to0$ limit as long as the particle densities that determine $F_{AB}$ in the limit become independent of whether particles $A$ and $B$ are massive or massless. For example $\rho_A$ and $\rho_B$ can be functions of the velocities $p_i/E_A$ and $p_i/E_B$ respectively.

First we consider the $\lambda_3=0$ theory, where the tree level diagram is just the quartic vertex. For our first example we simply set $F_{AB}$ to a constant $F$. Then the leading term in the $m/E\to0$ limit is
\begin{align}
\frac{dN}{d\Omega}=8  \lambda_4^{2}\frac{F}{s} \left(5 \cos \! \left(\theta \right)^{4}+2 \cos \! \left(\theta \right)^{2}+1\right).\label{e13}
\end{align}
For the second example we introduce an extra product of velocities, $F_{AB}=(p_i^2/E_A E_B)F$. Then the leading term in $m/E\to0$ limit is
\begin{align}
\frac{dN}{d\Omega}=8 \lambda_4^{2} \frac{F}{s} \left(35 \cos \! \left(\theta \right)^{4}+54 \cos \! \left(\theta \right)^{2}+15\right).
\end{align}
As a third example we remove the luminosity factor altogether and simply sum the cross sections. Then the result is
\begin{align}
\frac{dN}{d\Omega}=16  \lambda_4^{2}\frac{1}{s} \left(5 \cos \! \left(\theta \right)^{2}+1\right).
\end{align}

These results are all positive definite. Since the individual amplitudes grow like $s^2$, any term in the sum in (\ref{e10}) has a factor that grows like $(s/m^2)^4/s$. We instead see that intricate cancellations produce $1/s$ behaviour at large $s$. This behaviour is the same as for theories that have standard unitarity bounds. We see that the cancellations are quite robust, leaving only the angular dependence sensitive to the precise beam description.

We can also consider the same scattering experiment for the $\lambda_4=0$ theory. In this case each of the tree diagrams is a one-particle exchange diagram. Performing the sum over initial and final states as described above again gives a result that falls like $1/s$ at large $s$. The $\theta$ dependence now features a $1/\sin(\theta)^4$ pole behaviour for forward and backward scattering, due to the exchange of (near) massless particles. For the same three examples the $\theta$ dependence is again positive definite.
\begin{figure}[h]
\begin{center}
\includegraphics[width=.7\textwidth]{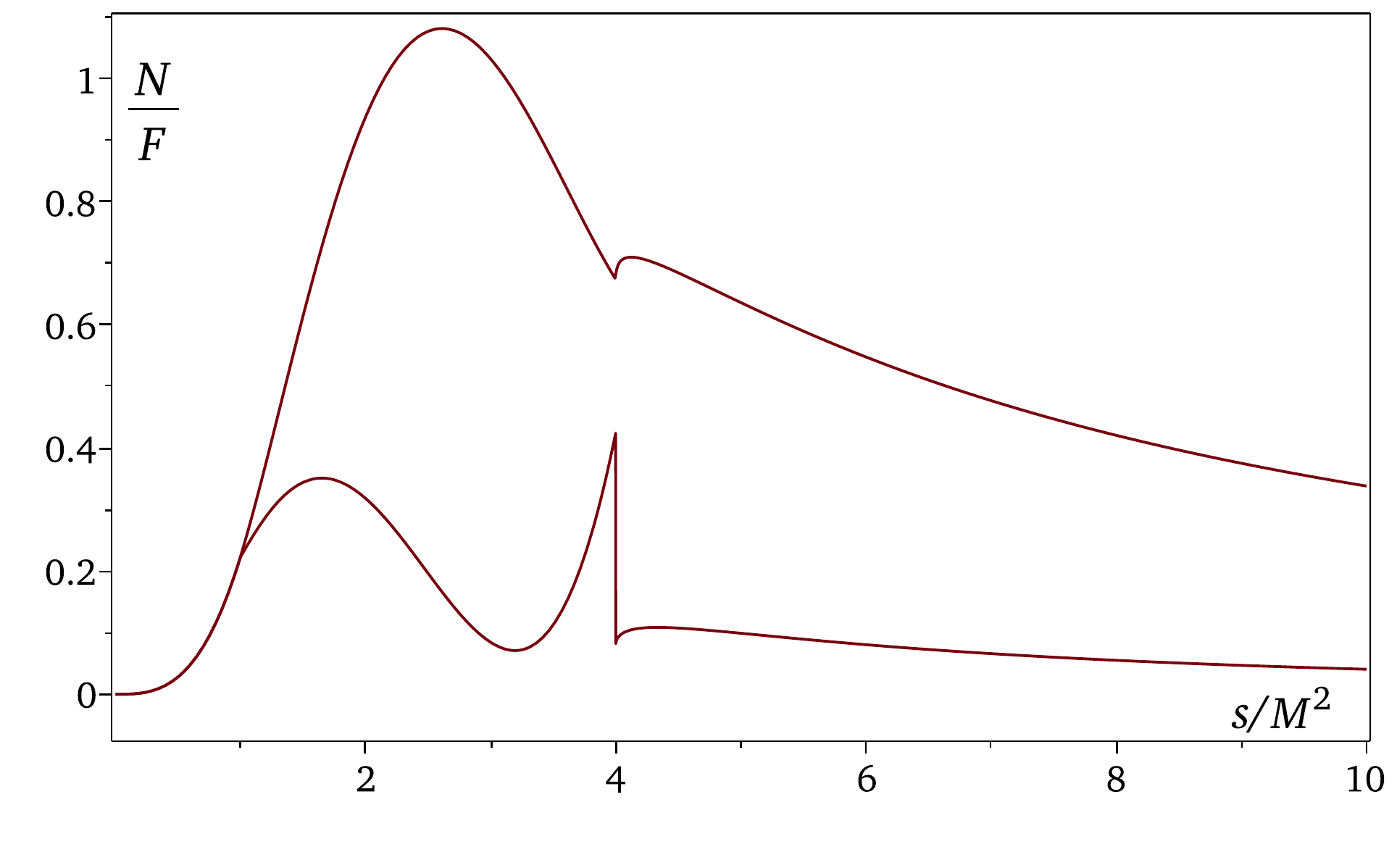}
\vspace{-2ex}
\caption{The number of scattering events according to (\ref{e10}) after the angular integration in the $\lambda_3=0$ theory for the choices of $F_{AB}=F$ (lower curve) and $F_{AB}=(p_i^2/E_A E_B)F$ (higher curve).}
\label{f1}
\end{center}
\end{figure}

Going back to the $\lambda_3=0$ theory, we can consider the number of scattering events in (\ref{e10}) without taking $m/E\to0$. After doing the angular integration we display the result in Fig.~\ref{f1} for the two different choices of $F_{AB}$, both of which give positive definite results. In the second case the extra factors of velocity suppress the $\psi_2$ contribution close to the thresholds where these massive particles are slow. The suppression of the negative norm contributions means that the final positive result is larger. Trouble finally arrives for the third example we considered, the naive sum over cross sections, which gives a result that is not positive definite. Here we notice the $1/p_i$ factor in (\ref{e12}), which is enhanced for a slow $\psi_2$ in the initial state. We have arrived back with the problem of an $n_2$-odd initial state that should not be treated in isolation.

\section{Comments}
We have presented a number of standard calculations within a 4-derivative quantum field theory with a negative norm state. The results show more similarities to standard 2-derivative theories than might have been expected. Rather paradoxically, it is the negative norm state itself that makes these similarities possible. The optical theorem being satisfied and the good high energy behaviour of inclusive scattering rates are both due to the cancellations brought about by the negative norm state. In addition we have a four dimensional scalar field theory that exhibits asymptotic freedom in the UV with a possible concurrent asymptotic freedom in the IR.  

One difference between 2 and 4 derivative theories involves the classical limit. The negative norm is a consequence of the manner in which the theory is quantized so as to ensure the propagation of positive energy only, and thus stability. There is no analog of negative norms in the classical theory, which instead directly exhibits negative energies and instabilities. We have mentioned that the theory becomes a free 2-derivative theory for wavelengths longer than $1/m$, and it is only here that the classical wave solutions make sense. The classical limit of quantum quadratic gravity must similarly involve long wavelengths where Einstein gravity applies. At wavelengths shorter than $1/m$ (or $1/m_{\rm Pl}$) these theories are intrinsically quantum.

Our approach to the theory has been strictly perturbative where, besides some unusual signs, standard Feynman methods apply. The perturbation theory displays a standard analytic structure.  An example of a deviation from this strictly perturbative approach is the resummation involving the one-loop self-energy diagram to obtain a dressed propagator. As soon as this is done the causal structure of the theory changes due to the abnormal structure of the dressed propagator. The dressed $\psi_2$ particle propagates backward-in-time over time scales of order its lifetime \cite{Grinstein:2008bg,Donoghue:2019ecz}. We have noted that a positive probability is associated with an $n_2$-even process, that is a process that involves both the appearance and disappearance of the $\psi_2$ particle.

At the higher energies and shorter times scales of interest to our study here, it is appropriate to use the strictly perturbative approach. We have applied this to the scattering within the $\psi_1$, $\psi_2$ sector, in particular for $m/E\to0$. A similar limit can be considered for the interactions between the $\psi_1$, $\psi_2$ fields and a normal matter sector. The couplings will involve the shift symmetric factor $\partial_\mu\phi=\partial_\mu(\psi_1-\psi_2)$ and so only this combination of positive and negative norm states can be produced. The cross section for some process that emits one $\phi$ can thus be calculated as a difference of two contributions that only differ by the appearance or not of $m$. The $1/m^2$ normalization factor is cancelled and the cross section remains finite as $m/M\to0$, where $M$ is some mass or energy scale characterizing the process. In addition the virtual exchange of the $\phi$ mode produces scattering cross sections among normal particles that behave sensibly as a function of energy. An example of this is photon scattering $\gamma\gamma\to\gamma\gamma$ at arbitrarily high energies in the context of quadratic gravity \cite{Holdom:2021oii}.

We end by considering a gauged extension of the theory. A $U(1)$ gauge field $A_\mu$ can be introduced via the Stuckelberg construction by defining $V_\mu=A_\mu+\partial_\mu\phi$. We also add the standard kinetic term for the photon,
\begin{align}
{\cal L}_1&=-\frac{1}{4g^2}F_{\mu\nu}F^{\mu\nu}+\frac{1}{2}V_\mu(\Box+m^2)V^\mu+\lambda_3V_\mu V^\mu\,\partial_\mu V^\mu+\lambda_4(V_\mu V^\mu)^2,\label{e23}\\
&F_{\mu\nu}=\partial_\mu V_\nu-\partial_\nu V_\mu=\partial_\mu A_\nu-\partial_\nu A_\mu.
\end{align}
The interest in this theory comes in the way it introduces cubic and quartic interactions among photons in a gauge invariant and renormalizable manner. In the appendix we give the $\beta$-functions of this theory.

\appendix*
\section{Running couplings in the gauged extension}
We see from (\ref{e23}) that the photon kinetic terms come both from the standard first term and from the $V^2$ term. We will continue to treat the renormalization of the $\frac{1}{2}V_\mu\Box V^\mu$ term as the wave function renormalization for the $\phi$ field, and thus also for the $A_\mu$ field. In turn this necessitates the independent renormalization of the standard first term and thus the gauge coupling $g$ appears in this term.

From the two types of $A_\mu$ kinetic terms, the photon propagator in Feynman gauge is
\begin{align}
\frac{-i\zeta\eta_{\mu\nu}}{k^2+i\epsilon},\quad \zeta=\left(1+\frac{1}{g^2}\right)^{-1}
.\end{align}
Our results will then have factors of $\zeta$ corresponding to photon propagators on internal lines.

The calculation of the renormalization constants will involve the same diagrams as before, with the addition of diagrams that involve photon propagators. We find
\begin{align}
Z_\phi Z_m-1&=-\left(\frac{3\lambda_4}{4\pi^2}+\frac{3\lambda_3^2}{8\pi^2}\zeta\right)\frac{1}{\varepsilon},\label{e30}\\
Z_\phi -1&=\left(\frac{5}{8}(1+\zeta^2)+\frac{1}{2}\zeta\right)\frac{\lambda_3^2}{\pi^2}\frac{1}{\varepsilon},\\
Z_\phi^\frac{3}{2} Z_3-1&=-\left(\frac{5}{4}+2\zeta+3\zeta^2\right)\frac{\lambda_4}{\pi^2}\frac{1}{\varepsilon},\\
Z_\phi^2 Z_4-1&=-\left(\frac{5}{4}+3\zeta+6\zeta^2\right)\frac{\lambda_4}{\pi^2}\frac{1}{\varepsilon},\\
Z_\phi Z_g^{-2}-1&=-\left(\frac{7}{12}(1+\zeta^2)+\frac{5}{12}\zeta\right)\frac{\lambda_3^2}{\pi^2}\frac{1}{\varepsilon}.
\end{align}

The new contribution in (\ref{e30}) still does not introduce scale dependence. Thus
\begin{align}
\gamma=-\gamma_m=\left(\frac{5}{16}(1+\zeta^2)+\frac{1}{4}\zeta\right)\frac{\lambda_3^2}{\pi^2}.
\end{align}
The $\beta$-functions are
\begin{align}
\frac{d\lambda_3}{d\ln\mu}&=-\left(\frac{5}{4}+2\zeta+3\zeta^2\right)\frac{\lambda_4\lambda_3}{\pi^2}-\frac{3}{4}\left(\frac{5}{4}(1+\zeta^2)+\zeta\right)\frac{\lambda_3^3}{\pi^2},\\\frac{d\lambda_4}{d\ln\mu}&=-\left(\frac{5}{4}+3\zeta+6\zeta^2\right)\frac{\lambda_4^2}{\pi^2}-\left(\frac{5}{4}(1+\zeta^2)+\zeta\right)\frac{\lambda_4\lambda_3^2}{\pi^2},\\\frac{dg}{d\ln\mu}&=\left(\frac{29}{48}(1+\zeta^2)+\frac{11}{24}\zeta\right)\frac{\lambda_3^2}{\pi^2}.
\end{align}
The runnings of $\lambda_3$ and $\lambda_4$ have all the same characteristics as before. The running of the $U(1)$ gauge coupling is not asymptotically free, as usual, and so this theory as a whole is not UV complete.


\end{document}